\documentclass[lettersize,hidelinks,journal]{IEEEtran}
\usepackage{cite}
\usepackage{amsmath,amssymb,amsfonts}
\usepackage{algorithmic}
\usepackage{graphicx}
\usepackage{textcomp}
\usepackage{amsmath,amsfonts,amssymb}
\usepackage{bm}
\usepackage[latin1]{inputenc}
\usepackage{algorithmic}
\usepackage{algorithm}
\usepackage{array}
\usepackage[caption=false,font=normalsize,labelfont=sf,textfont=sf]{subfig}
\usepackage{textcomp}
\usepackage{stfloats}
\usepackage{url}
\usepackage{verbatim}
\usepackage{graphicx}
\usepackage[acronym,shortcuts]{glossaries}
\usepackage{cite}
\usepackage{color}
\usepackage{lipsum}
\usepackage{url}
\usepackage{bm}
\usepackage{tablefootnote} 
\usepackage{orcidlink}

\newacronym{NG}{NG}{Next-Generation}
\newacronym{IoT}{IoT}{Internet of Things}
\newacronym{D2D}{D2D}{Device-to-Device}

\newacronym{NTT}{NTT}{Number Theoretical Transform}

\newacronym{QAM}{QAM}{Quadrature Amplitude Modulation}
\newacronym{4QAM}{4QAM}{4 Quadrature Amplitude Modulation}
\newacronym{MPSK}{MPSK}{M-ary Phase Shift Keying}

\newacronym{SNR}{SNR}{Signal to Noise Ratio}
\newacronym{SotA}{SotA}{State-of-the-Art}
\newacronym{P2P}{P2P}{Point-to-Point}
\newacronym{MIMO}{MIMO}{Multiple-Input Multiple-Output}

\newacronym{TDD}{TDD}{Time Divison Duplex}
\newacronym{CSI}{CSI}{Channel State Information}

\newacronym{AWGN}{AWGN}{Additive White Gaussian Noise}
\newacronym{BER}{BER}{Bit Error Rate}
\newacronym{BLE}{BLE}{Bluetooth Low Energy}
\newacronym{KER}{KER}{Key Error Rate}
\newacronym{SER}{SER}{Symbol Error Rate}
\newacronym{BCH}{BCH}{Bose-Chaudhuri-Hocquenghem}
\newacronym{CDF}{CDF}{Cumulative Distribution Function}

\newacronym{PLS}{PLS}{Physical Layer Security}

\newacronym{NIST}{NIST}{National Institute of Standards and Technology}
\newacronym{AES}{AES}{Advanced Encryption Standard}
\newacronym{RNG}{RNG}{Random Number Generator}

\newacronym{KEM}{KEM}{Key Encapsulation Mechanism}
\newacronym{ML-KEM}{ML-KEM}{Module-Lattice-based Key Encapsulation Mechanism}
\newacronym{PKE}{PKE}{Public Key Encryption}
\newacronym{PQ}{PQ}{Post-Quantum}
\newacronym{PQC}{PQC}{Post-Quantum Cryptography}

\newacronym{LWE}{LWE}{Learning With Errors}
\newacronym{MLWE}{MLWE}{Modular Learning With Errors}
\newacronym{MLWR}{MLWR}{Modular Learning With Rounding}
\newacronym{SVP}{SVP}{Shortest Vector Problem}
\newacronym{FO}{FO}{Fujisaki-Okamoto}

\newacronym{CPA}{CPA}{Chosen-Plaintext Attack}
\newacronym{CCA}{CCA}{Chosen Ciphertext attack}
\newacronym{IND}{IND}{Indistinguishability-under}
\newacronym{IND-CPA}{IND-CPA}{Indistinguishability under Chosen Plaintext Attacks}
\newacronym{IND-CCA}{IND-CCA}{Indistinguishability under Chosen Ciphertext Attacks}

\newacronym{WKyber}{WKyber}{Wireless Kyber}

\def\BibTeX{{\rm B\kern-.05em{\sc i\kern-.025em b}\kern-.08em
    T\kern-.1667em\lower.7ex\hbox{E}\kern-.125emX}}

\begin{document}

\title{Post-Quantum Wireless-based Key Encapsulation Mechanism via CRYSTALS-Kyber for Resource-Constrained Devices}

\author{M.A. Gonz\'alez de la Torre\textsuperscript{$\dag$\orcidlink{0000-0001-8398-1884}}, I.A. Morales Sandoval\textsuperscript{$\ddag$ \orcidlink{0000-0002-8601-5451}},\\
Giuseppe Thadeu Freitas de Abreu\textsuperscript{$\ddag$\orcidlink{0000-0002-5018-8174}}, L. Hern\'andez Encinas\textsuperscript{$\dag$\orcidlink{0000-0001-6980-2683}}
\thanks{$\dag$ M.A. Gonz\'alez de la Torre and L. Hern\'andez Encinas are with the Institute of Physical and Information Technologies (ITEFI) at Spanish National Research Council (CSIC), Madrid, Spain (e-mail: [ma.gonzalez, luis.h.encinas]@csic.es)}
\thanks{$\ddag$ I.A. Morales Sandoval and G.T. Freitas de Abreu are with the School of Computer Science and Engineering, Constructor University, Campus Ring 1, 28759 Bremen, Germany (e-mail: [imorales, gabreug]@constructor.university)}}

\maketitle

\begin{abstract}
We consider the problem of adapting a Post-Quantum cryptosystem to be used in resource-constrained devices, such as those typically used in Device-to-Device and
Internet of Things systems.
In particular, we propose leveraging the characteristics of wireless communications channels to minimize the complexity of implementation of a Post-Quantum public key encryption scheme, without diminishing
its security.
To that end, we focus on the adaptation of a well-known cryptosystem, namely CRYSTALS-Kyber, so as to enable its direct integration into the lowest layer of the communication stack, the physical
layer, defining two new transport schemes for CRYSTALS-Kyber to be used in Device-to-Device communications, both of which are modeled under a wireless channel subject to Additive
White Gaussian Noise, using a 4 Quadrature Amplitude Modulation constellation and a \acs{BCH}-code to communicate CRYSTALS-Kyber's polynomial coefficients.
Simulation results demonstrate the viability of the adapted Kyber algorithm due to its low key error probability, while  maintaining the security reductions of the original Kyber by considering the error distribution imposed by the channel on the cipher.
\end{abstract}


\glsresetall

\section{Introduction}
\vspace{1ex}

\IEEEPARstart{T}{oday}'s communication networks are characterized by an exponentially increasing number of wirelessly connected devices, which are expected to grow to up to 5 billion devices by the year 2025 \cite{ericsson19}, and which will be dominated by the deployment of \ac{NG}-\ac{IoT}.
A very important characteristic of this new development in wireless communications, is that the introduced devices will have a very diverse set of capabilities, ranging from computational, to power storage, to radio resources.
This poses new challenges to security, authentication and data privacy, which need to be provided for these massive networks of heterogeneous devices \cite{fang18, ahmad19}.

Concomitant with this trend, advances towards the widespread adoption of quantum computers are constantly being made \cite{deutsch1985quantum, cross2019validating, cumming2022using, ichikawa2023comprehensive}.
The use of \ac{PQ} \ac{PKE} algorithms, and more generally post-quantum cryptography, is part of the strategy to build barriers against the threats of cryptanalytic attacks implemented on quantum computers \cite{Shor:1997:PTA}.
In August 2024 the FIPS 203 was published by \ac{NIST}, establishing the first \ac{PQ} \ac{KEM} standard, known as \ac{ML-KEM} \cite{FIPS:2024:203}, which is based on the practically integral adaptation of CRYSTALS-Kyber \cite{NIST:3R:kyber, MaringerIFS2023}.

 Despite the individual security guarantees provided by each individual component in a system, it is well-known \cite{WangISPA2015,AndreaISCC2015,JungITJ2022} that their defenses can be compromised by targeting their weakest link.
In the context of wireless networks, these typically are \acs{IoT} devices with low power and computational capabilities which communicate in a \ac{D2D} fashion, which present strong challenge for the implementation of conventional \ac{PQ} encryption schemes, due to their resource constraints and strong reliance on features from upper layers of communication \cite{HuangIFS2024}.
In light of this, \ac{PLS} is a technology that can help alleviating the computational requirements at upper layers of communications \cite{porambage21, MucchiOJCS2021, AngueiraCST2022,katsuki2022noncoherent}.

In the context of this paper, \ac{PLS} is referred to as the body of practical mechanisms which leverage the characteristics of wireless communications media to increase a system's security. This focus is distinct from the pioneering work of Wyner\cite{wyner1975wire}, where the concept of perfect information-theoretical secrecy is introduced.
Instead, we address the integration of cryptography with features of \ac{PLS} and propose, in particular, a mechanism to incorporate the well-known cryptographic Kyber \ac{KEM} into a wireless communication system.  Similar efforts to implement \ac{PQ} security directly at the physical layer of communication systems have recently been published \cite{senlin24, ABDALLAH24}.
However, these \ac{SotA} approaches are not suitable for resource-constrained devices, as they require massive \ac{MIMO} capabilities, and modulation schemes with large constellation sizes.

Kyber is a public key \ac{PQ} algorithm, based on lattice problems that are currently considered hard, even against quantum attacks \cite{TosunIFS2024}.
The design of Kyber constitutes a \ac{KEM}, which is a public key algorithm specifically intended to perform key exchanges \cite{NosouhiIFS2023}, and an underlying \ac{PKE} that creates instances of the \ac{MLWE} problem as both public keys and ciphertexts.
Kyber encryption necessitates an error distribution to create instances of the \ac{MLWE} problem.
Consequently, a novel implementation of Kyber is proposed, incorporating an error by utilizing an \ac{AWGN} channel and a \ac{4QAM} constellation  \cite{MaringerIFS2023}.

The contributions of the article can be summarized as follows:
\begin{itemize}
\item A physical layer transport scheme for the polynomial coefficients of the CRYSTAL-Kyber cryptosystem is presented and modeled  for a wireless channel subjected to \ac{AWGN}, using a \ac{4QAM} constellation for symbol encoding and \ac{BCH} code for error correction.
\item Two adaptations of Kyber to the defined physical layer transport scheme are introduced as \ac{WKyber} V1 and V2. The first, \ac{WKyber} V1,  constitutes a \ac{PKE} and a KEM, which reduces the reliance on the original scheme's Binomial Distribution \ac{RNG} which is used for error sampling. The second, \ac{WKyber} V2, is a \ac{PKE}  scheme which further reduces the reliance on the aforementioned \ac{RNG} reliance by eliminating its use in other parts of the cryptosystem.
This is in contrast to the standard Kyber definition, which  necessitates error-free information transport which usually can only be achieved by implementing the upper layers of a communication system.
\item An analysis of the security of \ac{WKyber}, in particular how the modifications affect the security of the original cryptosystem and how the security of \ac{WKyber} can be estimated as a function of the parameters of the channel. This is achieved by considering the error distribution imposed by the transport scheme to the cipher's polynomial coefficients.
\end{itemize}

The remainder of this work is structured as follows.
The main properties, characteristics and definitions of the original CRYSTALS-Kyber system are described in Section \ref{S:CKyber}.
The physical layer transport scheme, channel model and encoding are described in Section \ref{S:Wireless}, together with the error probabilities for the transmitted polynomial coefficients.
The proposal of the two versions of \ac{WKyber}, along with the analysis of its security reductions, is presented in Section \ref{S:WKyber}. This section also comprises an analysis and comparison of the proposed Kyber modification under different
physical layer parameters. Additionally, the section provides an analysis of the implementation viability of the aforementioned modification.
Finally, in Section \ref{S:Conclusion} the conclusions are presented.

\noindent \textit{\textbf{Notation:}}
Column vectors and matrices are respectively denoted by lower- and upper-case bold face letters.
The transpose operation is indicated by the superscript $^\mathrm{T}$.
The operator $\leftarrow \mathcal{X}$ denotes sampling from a distribution $\mathcal{X}$, while, for any set $C$, $\leftarrow_R C$ denotes a uniformly random selection from $C$.
The inner product between two vectors is denoted by $\langle \bm{x}, \bm{y} \rangle$.
The $\lceil \cdot \rfloor$ operator denotes the rounding to the closest integer operation.

\vspace{-1ex}
\section{The CRYSTALS-Kyber cryptosystem}
\label{S:CKyber}

CRYSTALS-Kyber is a latticed-based \ac{PQ} cryptosystem proposed as the standard \ac{ML-KEM} by NIST in the summer of 2024 \cite{FIPS:2024:203}.
Kyber consists of two algorithms 1) Kyber \ac{PKE}, the Public Key Encryption algorithm, which provides the security,

 and 2) Kyber \ac{KEM}, the Key Encapsulation Mechanism that  defines the execution of the given \ac{PKE} to exchange  keys between users. The security features, key, ciphertext generation and plaintext recovery, of these algorithms all rely fundamentally on the \ac{LWE} problem \cite{Peikert:2009:PKC, RaviIFS2022}.
The rest of this section focuses on introducing these fundamental notions.

Let $L$ be a lattice, then the \ac{LWE} problem can be stated as follows: given pairs $(\bm{a}_i, b_i)$, such that $ \bm{a}_i\leftarrow_R {L}$ and $b_i = \langle \bm{s}, \bm{a}_i \rangle + e_i$, where $e_i\leftarrow \mathcal{X}$ is an error, the goal is to find the secret vector $\bm{s} \in L$. If no algebraic structure is considered on the lattice, then $L = \mathbb{Z}_{q}^{n}$.
 The objective of the problem is to determine the vector $\bm{s}$ from several samples as follows
\begin{subequations}
\begin{eqnarray}
&\bm{a}_1 \in \mathbb{Z}^{n}_{q}, \quad b_1 = \langle  \bm{s}, \bm{a}_1 \rangle + e_1,& \\
&\bm{a}_2 \in \mathbb{Z}^{n}_{q}, \quad b_2 = \langle  \bm{s}, \bm{a}_2 \rangle + e_2,& \\[-1ex]
&\vdots& \nonumber\\
&\bm{a}_k \in \mathbb{Z}^{n}_{q}, \quad b_r = \langle  \bm{s}, \bm{a}_k \rangle + e_k.&
\end{eqnarray}
\end{subequations}

In the case of Kyber the modular version of LWE  is considered, i.e. the chosen lattice is also a module for a polynomial ring.
Therefore, if $R_q := \mathbb{Z}_q[x]/(x^n+1)$ is the finite ring of the polynomials of degree less than $n$ with coefficients in $\mathbb{Z}_q$, the lattice considered in Kyber is $R_q^k = \left(\mathbb{Z}_q[x]/(x^n+1)\right)^k$.
The parameters $q$, $n$, and $k$ define the lattice, $R_q^k$, where $q$ denotes the modulus of the coefficients, $n$ is the degree of the polynomials of the ring $R_q$, and $k$ is the range of $R_{q}^{k}$ as a module.
Finally, parameters $\eta_1$ and $\eta_2$ define the range of the error distributions $\mathcal{X}_1 = \mathcal{B}_{\eta_1}$ and $\mathcal{X}_2 = \mathcal{B}_{\eta_2}$ respectively, where the  binomial distribution $\mathcal{B}_{\eta_i}$ is centered at 0 and has range $[-\eta_i, \eta_i]$, $i\in \{1,2\}$.

The Kyber cryptosystem defines two functions\cite{NIST:3R:kyber} that compress and decompress its inputs component-by-component, given respectively by
\begin{subequations}
\begin{equation}
\text{Compress}_q(x,d) \triangleq \bigg\lceil \frac{2^d}{q}\cdot x \bigg\rfloor \big(\mathrm{mod}^+ 2^d\big),
\label{eq:compress}
\end{equation}
\begin{equation}
\text{Decompress}_q(x,d) \triangleq \bigg\lceil \frac{q}{2^d}\cdot x \bigg\rfloor.
\label{eq:decompress}
\end{equation}
\end{subequations}

These two functions apply a quantification of the polynomial coefficients belonging to the $[0, q-1]$ interval and approximate them to the closest value of the $[0,2^d]$ interval. Applying the $\text{Compress}(\cdot)$ creates an instance of the \ac{MLWR} problem, which is a variant of \ac{MLWE} where the small error terms are already determined rather than sampled, and this error is avoided by rounding from one modulus to a smaller one. It is specified in \cite{NIST:3R:kyber} that the error introduced by the compress function is not considered in the security analysis.

The structure of Kyber \ac{PKE} is as follows:
\begin{enumerate}
\item The private and public keys are defined as $sk:=\bm{s}$ and $pk := (\bm{A}, \bm{b})$ respectively; where $\bm{A} \in R_q^{k \times k}$ is a pseudorandom matrix and $\bm{b} := \bm{A} \bm{s} + \bm{e}$; with $\bm{e}$, $\bm{s} \in R_{q}^{k}$, and $\bm{e}$, $\bm{s}\leftarrow \mathcal{X}_1$.

\item A binary message $m \leftarrow_R \{0,1\}^n$ is selected, and errors are sampled such that $\bm{s}' \leftarrow \mathcal{X}_1$ and  $\bm{e}'$, $e'' \leftarrow \mathcal{X}_2$, where $\bm{s}'$, $\bm{e}' \in R_q^k$ and $e'' \in R_q$.

\item The ciphertext $c$ is defined as
\begin{equation}
c := \big( \text{Compress}_q(\bm{u}, d_u), \text{Compress}_q(v, d_v) \big),
\end{equation}
with $\bm{u} =  \bm{A}^\mathrm{T} \bm{s}' + \bm{e}'$, $v =  \bm{b}^\mathrm{T} \bm{s}' + e'' + \hat{m}$, and $\hat{m} = \text{Decompress}_q(m, 1)$.

\item The decryption of the message is then given by
\begin{align}
v - \bm{s}^\mathrm{T} \bm{u} & =  \bm{b}^\mathrm{T} \bm{s}' + e'' + \hat{m}  - \bm{s}^\mathrm{T} \! \bm{A}^\mathrm{T} \! \bm{s}' - \bm{s}^T \bm{e}' \nonumber \\
& = \bm{e}^\mathrm{T} \bm{s}' \! + \bm{s}^\mathrm{T} \! \bm{A}^\mathrm{T} \! \bm{s}' \! + e'' \! + \hat{m} \! - \bm{s}^\mathrm{T} \! \bm{A}^T \! \bm{s}' \! - \bm{s}^\mathrm{T} \bm{e}' \nonumber \\
& =  \hat{m} + \bm{e}^\mathrm{T} \bm{s}' - \bm{s}^\mathrm{T} \bm{e}' + e''.
\end{align}

\item The original binary message can then be recovered:
\begin{equation}
m = \text{Compress}_q(\hat{m} + \bm{e}^\mathrm{T} \bm{s}' - \bm{s}^\mathrm{T} \bm{e}' + e'', 1).
\end{equation}
\end{enumerate}

In the absence of the secret $\bm{s}$, any party trying to access the message $m$ must solve the \ac{LWE} instance that presents any ciphertext or public key. The level of security provided by this problem depends on the set of parameters $q$, $n$, $k$, $\eta_1$, and $\eta_2$, which have previously been introduced in the description of the structure of the \ac{LWE}-based \ac{PKE} system.

\begin{table}[H]
\begin{center}
\caption{Sets of parameters for Kyber}
\label{gonza.t1}
\vspace{-1ex}
\begin{tabular}{|@{\;}c@{\;}|@{\;}c@{\;}|@{\;}c@{\;}|@{\;}c@{\;}|@{\;}c@{\;}|@{\;}c@{\;}|@{\;}c@{\;}|@{\;}c@{\;}|} \hline
Parameters  & \ac{NIST}-level & $q$ & $n$ & $k$ & $\eta_1$ & $\eta_2$ & $(d_u,d_v)$ \\ \hline
Kyber512    & 1 & $3329$ & $256$ & $2$ & $3$ & $2$ & $(10, 4)$ \\ \hline
Kyber768    & 3 & $3329$ & $256$ & $3$ & $2$ & $2$ & $(10, 4)$ \\ \hline
Kyber1024   & 5 & $3329$ & $256$ & $4$ & $2$ & $2$ & $(11, 5)$ \\ \hline
\end{tabular}
\end{center}
\vspace{-3ex}
\end{table}

In its final release, three sets of parameters were defined for Kyber and can be found in Table~\ref{gonza.t1}.
Kyber512, 768 and 1024 target security levels 1, 3, and 5 established by \ac{NIST}.
These security levels were defined as follows: Any attack that breaks a security definition, must require computational resources comparable to or greater than those required for a search on a blockcipher of \ac{AES} with either 128-, 192-, or 256-bit keys respectively (i.e. \ac{AES}128, \ac{AES}192 or \ac{AES}256).
In this work we focus in the parameter set for Kyber768, that corresponds to the security level 3.

Kyber is a cryptosystem with \ac{IND-CCA} semantic security. In particular, Kyber \ac{PKE} reaches \ac{IND-CPA} security, while Kyber \ac{KEM} reduces the \ac{IND-CCA} security to the \ac{IND-CPA} of Kyber \ac{PKE}. IND-CPA security is defined as the probability of success of an attacker that has to choose between two messages given the ciphertext of one of them, chosen randomly. If the adversary has access to a decryption oracle, then it is considered \ac{IND-CCA} security \cite{Jiang:2018:FO}.

\section{Wireless Channel Physical Layer}
\label{S:Wireless}

\subsection{System Model}\label{ss:SysModel}

The public communication medium is modeled as a memoryless wireless communication channel with \ac{AWGN} and no fading. The communication takes place between two independent devices, the complex baseband received signal is then described as \cite{tse}
\vspace{-0.5ex}
\begin{equation}
\mathsf{y} = s + n,
\vspace{-0.5ex}
\end{equation}
where $s \in \mathbb{C}$ is the complex transmitted symbol, $n \in \mathbb{C} \sim \mathcal{N}(0,\sigma)$ is the circularly symmetric Gaussian noise added by the wireless channel, and $y \in \mathbb{C}$ is the received symbol.

 The assumption of a Gaussian channel is made here only to simplify the error calculations used for security, without fundamentally compromising the security guarantees of the scheme, which hold also under fading channel models.
In particular, fading channels with perfect channel estimation at the receiver only affect the bit error rates in proportion to the fading rate, with a further degradation occurring under imperfect channel knowledge.
In either of these cases, the higher \ac{BER} needs merely be compensated, either by requiring a higher SNR, or by introducing diversity \cite{DumanBookCh2007} or coding techniques for fading channels \cite{BIGLIERI20001135}, so as to keep the key-agreement rates.

For the sake of illustration, the transmission of \ac{4QAM} symbols is considered, each representing a 2-bit word in a 4 symbol alphabet, with in-phase and quadrature components given by
\begin{subequations}
\begin{equation}
A_i = \pm \sqrt{E_b},
\end{equation}
\begin{equation}
A_q = \pm \sqrt{E_b},
\end{equation}
where $E_b$ is the energy per bit and each of the 4 transmitted symbols is described as
\begin{equation}
s_n = A_i \mathcal{I} + A_q \mathcal{Q}, \quad \forall n \in \{0,1,2,3\}.
\end{equation}
\end{subequations}

It is also assumed that a Gray code is employed over the constellation, such that the Hamming distance between any two adjacent symbols is minimal and equal to one.
Under this modulation scheme, a data block is mapped into a string of symbols to be transmitted over the wireless communication channel.

Finally, the received baseband symbols, $s$, are decoded back into data by observing in which one of the four quadrants they lie. Using this decoding scheme, the \ac{BER} approximation for a Gray-coded \ac{4QAM} constellation transmitted over an AWGN channel \cite{tse} is given by the bit error probability
\begin{equation}
P_b \approx \mathcal{Q}\Bigg( \sqrt{2\frac{E_b}{N_0}} \Bigg),
\label{eq:p_4QAM}
\end{equation}
where $E_b/N_0$ is the ratio of energy per bit to the noise power spectra density given in linear form, and $\mathcal{Q}(x)$ is the Q-function given by \cite{Proakis:2007:DCom}

\vspace{-1ex}
\begin{equation}
\label{eq:BER}
\mathcal{Q}(x) = \frac{1}{\sqrt{2\pi}} \int^{\infty}_x \exp \Big(-\frac{t^2}{2}\Big) dt.
\end{equation}

The behavior of the error probability\footnote{This formulation is only valid at high signal to noise ratios} is described by Eq.~\eqref{eq:BER}, where $x$ is a \ac{SNR}, which in turn can be also represented in terms of energy-per-bit over noise ($E_b/N_0$) via the relation
\begin{equation}
\text{SNR}  = 10 \log_{10} (E_b/N_0).
\end{equation}

\subsection{WKyber Transport Protocol}

The objective of the physical layer employed in this scheme is to bring the exchanges required to run CRYSTALS-Kyber down to the signal level of information exchange, without the use of upper layer protocols.
For the sake of clarity, we describe in the sequel how the information is transported down to the symbol level during an exchange.
It is important to note that under the assumption of AWGN, the error probability of each transmitted symbol/bit is independent and identically distributed, which allows us to more easily calculate the error distribution for the encoded transmitted words.

From the beginning, it was clear that naively encoding the coefficients of the polynomials as single symbols is not viable, as the constellation size required is that of $q=3329$ elements.
It is known \cite{tse} that the bit error probability for an MPSK (M-ary Phase Shift Keying) constellation is

\begin{equation}
P_b \approx \frac{2}{\log_2 M} \mathcal{Q}\bigg( \sqrt{2\frac{E_b}{N_0} \log_2 M} \sin\Big(\frac{\pi}{M}\Big) \bigg),
\end{equation}
where $M$ is the number of elements in the constellation.

In other words, prohibitively large transmit signal powers would be necessary to achieve  the low error rates required to make this transmission scheme viable.

\begin{figure}[h]
\centering
\includegraphics[width=\columnwidth]{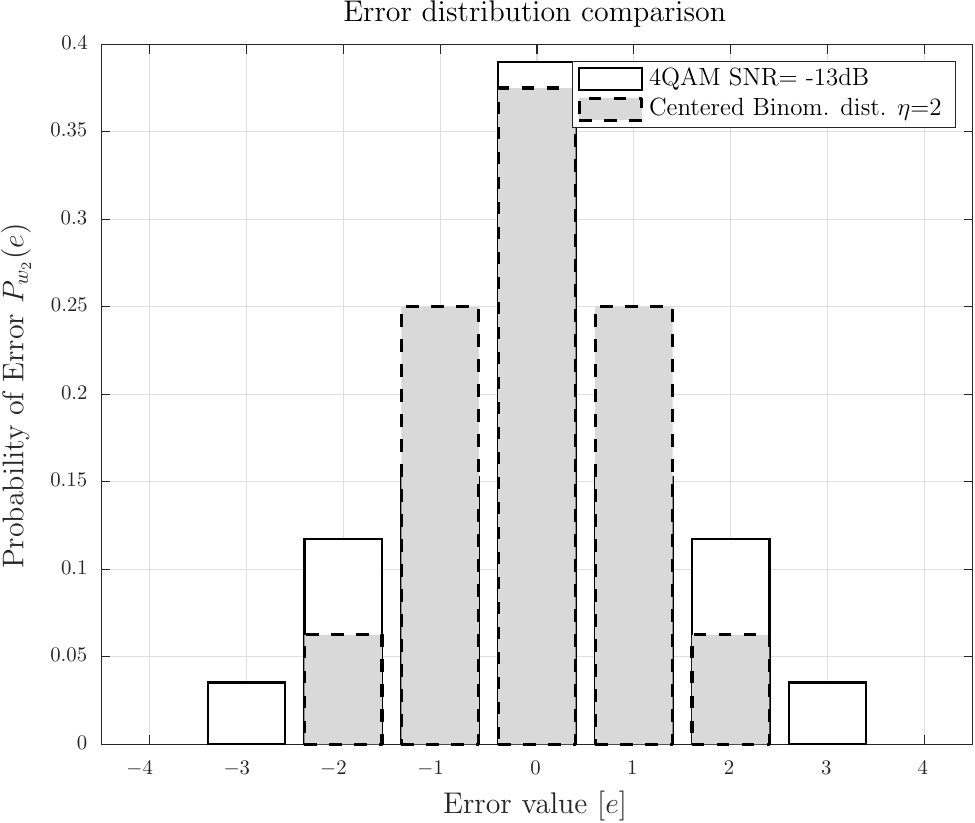}
\caption{2 Bit Error Probability for -13~dB \ac{SNR}}
\label{fig:2bit_binom}
\vspace{3ex}
\includegraphics[width=\columnwidth]{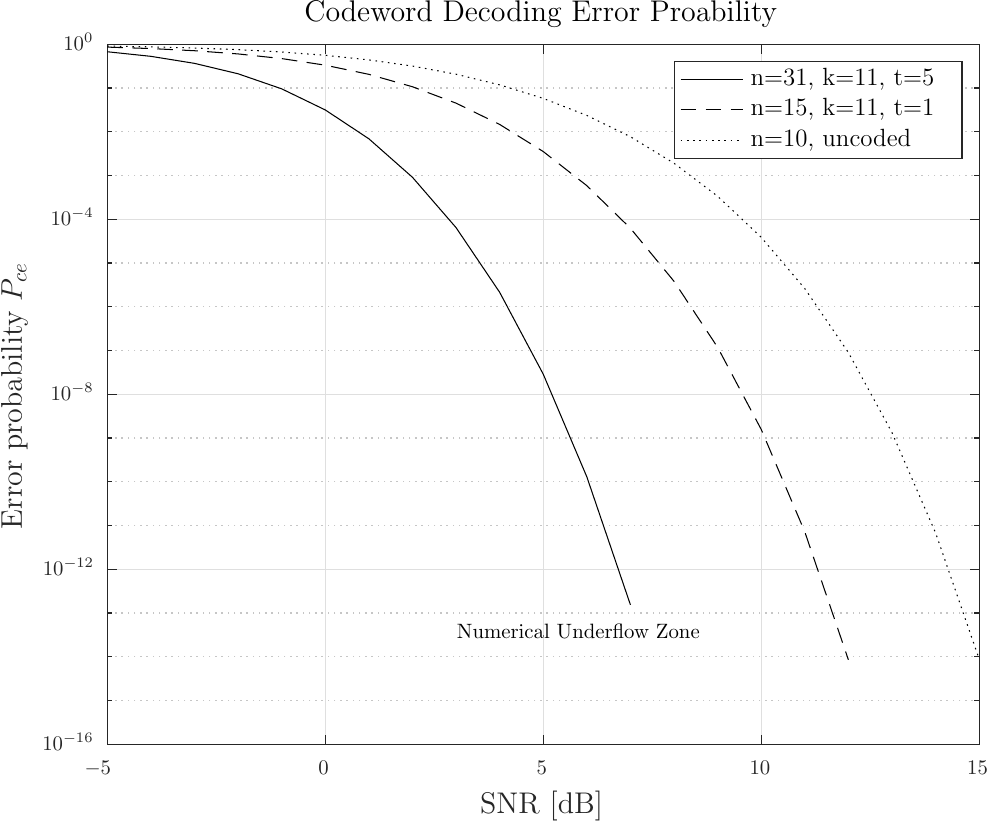}
\caption{Codeword error decoding probability against SNR}
\label{fig:Codeword}
\end{figure}

In order to circumvent this challenge, the symbol constellation presented in Section~\ref{ss:SysModel} was used.
However, this now requires to encode the polynomial coefficients in $\mathbb{Z}_{3329}$ into a string of 2-bit symbols before transmission, thus fitting into a 6-symbol/12-bit word.
In the original Kyber scheme, the $u$ and $v$ values are mapped from $\mathbb{Z}_q$ to $\mathbb{Z}_{2^{d}}$ values, followed by a bit packing step which moved the coefficients into $d$ bit words.
However, we instead skip the compression step for all coefficients and separate each 12 bit word into the 10 most significant bits, and the 2 least significant bits
\begin{equation}
w = \underbrace{b_{11} b_{10} b_9 b_8 b_7 b_6 b_5 b_4 b_3 b_2}_{w_{10}}\underbrace{b_1 b_0}_{w_2}.
\vspace{-0.5ex}
\end{equation}

This packing is specially problematic as all symbols experience the same bit error probability regardless of their significance within the word, resulting on equal likelihoods for a coefficient to have a shift of $\pm2048$ units as that of $\pm1$ unit in value.

To protect against this scenario, the upper word $w_{10}$ is encoded using a \ac{BCH} code, and the lower word $w_2$ is then mapped into a single \ac{4QAM} symbol.
For now, assuming a sufficiently high protection to the most significant bits of the coefficient from the \ac{BCH} code, the worst case scenario for the difference in magnitude between transmitted and received coefficients is equal to 3.
These transitions are represented by
\begin{equation}
(11)_2 \rightarrow (00)_2, \text{with } e = -3,
\vspace{-0.5ex}
\end{equation}
and
\begin{equation}
(00)_2 \rightarrow (11)_2, \text{with } e = 3.
\vspace{-0.5ex}
\end{equation}

Taking into account all potential error transitions in the last two bits $w_2$, the error distribution on each of the $[0,3328]$ information chunks representing each of the polynomial coefficients in $\mathbb{Z}_{3329}$ is independent and identically distributed and is approximately given by
\begin{equation}
\label{eq:pe_coeff}
P_{w_2}(e)=
\begin{cases}
0, &\text{if} \, |e| \geq 4\\
\frac{1}{4} P_b^2, &\text{if} \, |e| = 3\\
\frac{1}{2} P_b \cdot (1 - P_b), &\text{if} \, |e| = 2\\
\frac{1}{2} P_b \cdot (1 - P_b) + P_b^2, &\text{if} \, |e| = 1\\
\frac{1}{4} (1-P_b)^2, &\text{if} \, e = 0
\end{cases}
\end{equation}
where $P_b$ is calculated as a function of the \ac{SNR} as in Eq.~\eqref{eq:p_4QAM}.

This distribution is compared against the Kyber768's error distribution in Figure~\ref{fig:2bit_binom}.
In order to ensure the correctness of the approximation in Eq.~(\ref{eq:pe_coeff}), the probability of error of the upper 10-bits, $w_{10}$, must be sufficiently small, which can be achieved using a binary \ac{BCH} code to encode the 10 most relevant bits ($w_{10}$).
\ac{BCH} codes \cite{bc60,h59} are a class of cyclic error-correcting codes which are constructed using polynomials over a Galois field\cite{lin_error} with the following parameters\footnotemark\, for the binary case
\begin{IEEEeqnarray*}{L"L}
\text{Block length:}&       n = 2^m - 1, \\
\text{Parity check bits}&   n - k \leq mt,\\
\text{Minimum distance:}&   d_{min} \geq 2t + 1,
\end{IEEEeqnarray*}
where $m$ is the degree of the generator polynomial, $k$ is the number of information bits to encode, $t$ is the error correction capability of the code, and $d_{min}$ is the minimum distance between any two codewords.

\footnotetext{A \ac{BCH} code will exist for any case where $m \geq 3,\ t < 2^{m-1}$ and $m, t \in \mathbb{Z}$}

A validly constructed \ac{BCH} code will always be able to correct any pattern of $t$ or fewer errors in a block of $n$ digits\cite{lin_error}, as well as coming in a wide range of values for $n$ and~$k$.
The aforementioned versatility in these well-known codes is why a \ac{BCH} code of length $n=31$ bits, error correction capability of $t=5$ bits over a message of at most $k=11$ bits has been chosen to protect the upper 10 bits, $w_{10}$, of the polynomial coefficient during transmission.

One could have potentially used a longer code to further minimize the error probability; however, this was not deemed necessary for an initial analysis due to the already minimal codeword decoding error probability shown in Figure~\ref{fig:Codeword}.

The probability that a transmitted codeword cannot be decoded can be easily calculated starting from the fact that the probability of error of each individual bit is known, independent and identically distributed. Therefore, each number of errors is characterized by a binomial distribution per its definition, and the probability of seeing between $0$ and $t$ errors is described by the \ac{CDF} $F(k;n,p)$ of a binomial distribution given by
\begin{subequations}
\begin{equation}
 F(k;n,p) = I_{1-p}(n-k, k + 1),
\label{eq:CDF-function}
\end{equation}
\begin{equation}
I_x(a,b) = \frac{B(x;a,b)}{B(a,b)},
\end{equation}
\begin{equation}
B(x;a,b) = \int^{x}_0 t^{a-1} (1-t)^{b-1} dt,
\end{equation}
\end{subequations}

\noindent where, for our case, we want the complement of this \ac{CDF}, finally yielding the probability to fail to recover a coded word
\begin{equation}
P_{ce} = 1 - I_{1-P_b}(n-t, t + 1),
\label{eq:Pce}
\end{equation}
where $P_b$ is the bit error probability of the underlying transmission scheme.

\section{Wireless Kyber}
\label{S:WKyber}

This section presents \acrfull{WKyber}, our work's primary contribution, which adapts CRYSTALS-Kyber's algorithms and message exchange protocols to maximize the usage of a wireless channel's physical layer security properties.
Depending on the degree of embedding of WKyber into the physical layer of communications it can be separated into two different schemes, or "Versions".
Their high-level differences are summarized in Table~\ref{tb:WKyber-versions} and are differentiated by how many of Kyber's \ac{LWE} error instances are replaced by the \ac{AWGN} introduced by the wireless propagation media during the encryption procedure, i.e. Kyber \ac{PKE}.

In addition to the novel source of errors for the \ac{LWE} problem, changes to the compression scheme, ciphertext and public key (in the case of \ac{WKyber} V2) definitions were also required.
 Furthermore, it is crucial to ensure that the proposed scheme is secure, and for this reason a security and error probability analysis has been conducted.

\subsection{WKyber PKE}
\label{ss:WKyber-PKE}

In this subsection two proposals to adapt Kyber \ac{PKE} for the wireless channel are introduced as \ac{WKyber} V1 and V2.
They are key exchange algorithms capable of leveraging the properties of the \ac{AWGN} channel, designed to be analogous to the original Kyber \ac{PKE} system, to ensure that the \ac{FO} transformation \cite{Jiang:2018:FO}  can still be applied to the first version.
This results in an adapted yet recognizable \ac{KEM} scheme, where the distinction between two versions of the precursor \ac{PKE} depends on the depth of integration of the the public key into the physical layer of communication.

\begin{table}[h]
\begin{center}
\caption{Proposed versions of \ac{WKyber}}
\vspace{-1ex}
\label{tb:WKyber-versions}
\begin{tabular}{|l|l|c|c|} \hline
Version & Channel use & Scheme & Security \\ \hline \hline
V1 & ciphertext & \ac{PKE} and \ac{KEM} & \ac{IND-CCA}  \\ \hline
V2 & ciphertext and public key  & \ac{PKE}  & \ac{IND-CPA} \\ \hline
\end{tabular}
\end{center}
\vspace{-4ex}
\end{table}

\begin{table}[h]
\begin{center}
\caption{\ac{WKyber} \ac{PKE} V1: KeyGen, Encryption, and Decryption} \label{tb:WKyber-V1}
\resizebox{9cm}{!} {
\begin{tabular}{|lcr|} \hline
\multicolumn{3}{|c|}{\emph{WKyber \ac{PKE} V1}}\\ \hline \hline
{\bf KeyGeneration}() & & \\
Get $seed_A$ if not given &  &  \\
$\bm{A} = GenMatrix(seed_A)$ &  & \\
$\bm{s} \leftarrow \mathcal{B}_{\eta_1}$ &  & \\
$\bm{e} \leftarrow \mathcal{B}_{\eta_1}$ &  & \\
$\bm{b} = \bm{A} \bm{s} + \bm{e}$ &&  \\
$sk := \bm{s}$ && \\
$pk := (seed_A, \bm{b})$ &&\\
return $(pk,sk)$ &&\\
& $\stackrel{\text{AWGN}(pk, 10, 10)}{\longrightarrow}$ & \\
&& {\bf Encrypt$(pk, m; r)$} \\
&&  $\bm{A} = GenMatrix(seed_A)$  \\
&&  $\bm{s}' \leftarrow \mathcal{B}_{\eta_1}$ \\
&&  $\bm{u} = \bm{A}^\mathrm{T} \bm{s}'$  \\
&&  $v = \bm{b}^\mathrm{T} \bm{s}'  + \hat{m}$ \\
&& return $c := (\bm{u}, v)$  \\
& $\stackrel{\text{AWGN}(c, 10, -10)}{\longleftarrow}$ & \\
{\bf Decrypt}$(sk, c)$ & & \\
$m = \text{Compress}_q(v - \bm{s}^\mathrm{T} \bm{u}, 1)$ &&  \\
return $m$ & & \\\hline
\end{tabular}
}
\end{center}
\vspace{2ex}
\begin{center}
\caption{\ac{WKyber} \ac{PKE} V2: KeyGen, Encryption, and Decryption} \label{tb:WKyber-V2}
\resizebox{9cm}{!} {
\begin{tabular}{|lcr|} \hline
\multicolumn{3}{|c|}{\emph{WKyber \ac{PKE} V2}}\\ \hline \hline
{\bf KeyGeneration}() && \\
Get $seed_A$ if not given && \\
$\bm{A} = GenMatrix(seed_A)$ && \\
$\bm{s} \leftarrow \mathcal{B}_{\eta_1}$ && \\
$\bm{b} = \bm{A} \bm{s}$ &&  \\
$sk := \bm{s}$ && \\
$pk := (seed_A, \bm{b})$ &&\\
return $(pk,sk)$ && \\
& $\stackrel{\text{AWGN}(As, 10, -10)}{\longrightarrow}$ & \\
&& {\bf Encrypt}$(pk, m; r)$  \\
&& $\bm{A} = GenMatrix(seed_A)$ \\
&& $\bm{s}' \leftarrow \mathcal{B}_{\eta_1}$ \\
&& $\bm{u} = \bm{A}^\mathrm{T} \bm{s}'$ \\
&& $v = \bm{b}^\mathrm{T} \bm{s}' + \hat{m}$ \\
&&  return $c := (\bm{u}, v)$ \\
& $\stackrel{\text{AWGN}(c, 10, -10)}{\longleftarrow}$ & \\
{\bf Decrypt}$(sk, c)$ & & \\
$m = \text{Compress}_q(v - \bm{s}^\mathrm{T} \bm{u}, 1)$ && \\
return $m$ && \\ \hline
\end{tabular}
}
\end{center}
\vspace{-1ex}
\end{table}

\ac{PKE} V1 is presented in Table~\ref{tb:WKyber-V1}; this scheme limits its use of the \ac{AWGN} channel as a source of errors for ciphertext generation, while the public key generation and transmission is identical to the original Kyber.
The original Kyber submission used the \ac{FO}$^{\not\perp}$ transformation to build Kyber \ac{KEM} from Kyber \ac{PKE} because in \cite{Hofheinz:2017:MAF} it was proven that as long as Kyber \ac{PKE} is \ac{IND-CPA} secure, then Kyber \ac{KEM} will be \ac{IND-CCA} secure.
To reach this level of security, the \ac{FO}$^{\not\perp}$ transformation requires a re-encryption step inside the decapsulation algorithm, i.e. after the ciphertext is decrypted the resulting plaintext is encrypted again to verify if the new ciphertext matches one.
Utilizing \ac{WKyber} \ac{PKE} V1 allows for re-encryption, since the public key $(seed_A,\ \bm{b} = \bm{A}\bm{s} + \bm{e})$ does not depend on the uncontrollable conditions originating from the channel's effects. Therefore, if the security of \ac{WKyber} \ac{PKE} V1 is equivalent to that of the original Kyber, it is possible to apply the \ac{FO}$^{\not\perp}$ transformation and thus construct a \ac{KEM}, denoted as \ac{WKyber} {KEM} V1, of the same form.

When both the public key and ciphertexts are under the influence of \ac{AWGN} noise from the wireless channel, as it is for \ac{WKyber} \ac{PKE} V2, the original Kyber scheme requires further modifications than the ones needed for \ac{WKyber} \ac{PKE} V1.
Specifically, the \ac{LWE} instance of public key is redefined to be $A\cdot s$ and the receiver gets $pk = (seed_A, A\cdot s + e_{AWGN})$, where it and the ciphertext observe a low \ac{SNR} of $-10$~dB.
Therefore, the \ac{FO} transformation is not applicable anymore, as the sender of a public key of \ac{WKyber} \ac{PKE} V2 does not acquire a copy of it; and by definition is unable to replicate the encryption process.
The \ac{NIST} asserts that if a cryptosystem is \ac{IND-CPA} secure, its utilization with ephemeral keys, i.e. the generation of a new key pair at the beginning of each communication, becomes a viable proposition. Under this setup, \ac{WKyber} \ac{PKE} V2 is recommended as \ac{KEM} without re-encryption and utilizing ephemeral keys.

The proposed schemes are viable adaptations of the original Kyber system, each with its respective tradeoffs in the context of resource constrained devices. \ac{WKyber} \ac{KEM} V1 has the advantage that it reaches the higher level of \ac{IND-CCA} security, and thus allows the sender to re-use its public key. However, this comes at the cost of an usage of additional error correction. While \ac{WKyber} \ac{PKE} V2 relaxes both of these requirements by 1) leveraging the AWGN channel to introduce errors into the public key and 2) removing the need to provide error correction for the least significant bits of the public key.

Both versions of the construction of Wireless Kyber are compatible with all three Kyber parameters sets presented in Table~\ref{gonza.t1}.
These parameter sets differentiate only in the value of $k\in\{2, 3, 4\}$, and $k$ remains unaffected by the changes on the error distribution.
The primary focus is on the Kyber768 parameter set, as it is labeled as \emph{recommended}, while Kyber512 is labeled as \emph{light} and Kyber1024 as \emph{paranoid}.

\subsection{Other Modifications}
\label{ss:pack-and-cod}

Equations \eqref{eq:compress} and \eqref{eq:decompress} show the compression and decompression functions employed in the Kyber cryptosystem.
During the execution of a key exchange of Kyber, the $\text{Compress}_q$ function is utilized to compress the ciphertext during encapsulation.
This function is also used during decryption to transform the plaintext from a polynomial back to a bit string, and to erase the error added during the whole key exchange process.
It is important to differentiate between these two applications.
While the utilization of $\text{Compress}_q$ during the decryption is required for this process to work, the compression of the ciphertext is only considered to enhance the performance.
The adaptations introduced in \ac{WKyber} have an negative impact on the applicability of $\text{Compress}_q$.

A ciphertext of Kyber is conformed of a vector of polynomials $u$ and a polynomial $v$, with coefficients in the finite field $\mathbb{Z}_q$ with $q = 3329$. The $\text{Compress}_q$ function reduces the size of each coefficient from module $q$ to the parameters $(d_u,d_v) = (10,4)$ (see Table~\ref{gonza.t1}). The bit reduction applied by the compression function includes a few steps described as follows. If $d_u = 10$, the interval $[0, q-1]$ is quantified into $2^{10}$ sub-intervals, and subsequently, each integer in $[0, q-1]$ is rounded to the nearest division. Given that the alphabet left has $2^{10}$ total words, it can be expressed as $10$-bits words, ordered by size.

The compression of Kyber supposes adding an error to each coefficient of any ciphertext. This approach is deemed feasible due to the fact that the errors introduced are eliminated at the conclusion of the decryption process. In the case of \ac{WKyber} PKE, none of the proposed versions is compatible with this function, since its design relies on the transmission of the whole 12 bits of each coefficient of the ciphertext. In order to apply the \ac{BCH} code and recover the initial $10$ bits as well as to send the last two bits with a lower \ac{SNR}, and add the error in this way, it is necessary to send $12$ bits in total. Consequently, the $\text{Compress}_q$ function does not work with the introduced modifications.

The error added by the $\text{Compress}_q$ function is not considered in the security analysis, therefore not using this function does not affect the assumption that the \ac{WKyber} scheme as a whole resembles a standard Kyber instance with a different error distribution. However, it is acknowledged that the error generated by the compression function is considered in the calculation of the error probability of the cryptosystem. Thus, if this error is eliminated, the probability error during a key exchange is reduced.

\begin{subequations}
The adaptation of Kyber to this wireless implementation requires a new definition of the ciphertext, using the noise of the wireless communications channel to generate the errors. In the case of \ac{WKyber} \ac{PKE} V1, $\bm{e}'$ and $e''$ are generated by the channel, as well as the error $e$ in the key generation algorithm for \ac{WKyber} \ac{PKE} V2. Instead of sending the ciphertext composed of  $(\mathbf{u},v)$ where

\begin{equation}
\bm{u} \! = \! \text{Compress}_q \left( \bm{A}^\mathrm{T} \bm{s}' \! + \! \bm{e}', d_u \right),
\label{eq:uComp}
\end{equation}
\begin{equation}
v \! = \! \text{Compress}_q \left( \bm{b}^\mathrm{T} \bm{s}' \! + \! e'' \! + \! \text{Decompress}_q(m, 1), d_v \right),
\label{eq:vComp}
\end{equation}
\end{subequations}
the ciphertext is sent without adding an error sampled from the $\mathcal{B}_{\eta_i}$ distribution.

Consequently, the ciphertext in both \ac{WKyber} versions consists in the pair $(\mathbf{u}_{WK}, w_{WK})$, where
\begin{subequations}
\begin{equation}
\bm{u}_\text{WK} = \bm{A}^\mathrm{T} \bm{s}' ,
\label{eq:u_tx}
\end{equation}
\begin{equation}
v_\text{WK} = \bm{b}^T \bm{s}' + \text{Decompress}_q(m, 1).
\label{eq:v_tx}
\vspace{-2ex}
\end{equation}
\label{eq:cypher_tx}
\end{subequations}

In the case of \ac{WKyber} V2, a similar modification is also applied to the public key. In previous versions of the standard Kyber, the public key had a compression, but this is removed in the final draft of the cryptosystem \cite{Nist:2023:Draft1}, and in the final \ac{NIST} standard draft of \ac{ML-KEM} \cite{FIPS:2024:203}. The public key of Kyber is defined as $pk_{Kyber}=(seed_A, \bm{b} := \bm{A} \bm{s} + \bm{e})$. In the case of \ac{WKyber} V2 the error is also added through the channel, hence the information sent is $pk_{\ac{WKyber}V2}=(seed_A,\ \bm{A}\bm{s})$.

\subsection{Security and Error Distribution}
\label{ss:security}

In this section, a detailed analysis of the security of both versions of the WKyber cryptosystem is presented.
The fundamental argument supporting the security of WKyber is based on the premise that Kyber is regarded as a secure cryptosystem and that the modifications introduced do not compromise the original security.

The evaluation of \ac{PQC} cryptosystems is typically conducted considering three primary aspects: semantic security, the underlying hardness of the mathematical problem, and the security of the implementations.
The \ac{FO}$^{\not\perp}$ transformation applied in the original scheme is also considered for \ac{WKyber} \ac{KEM} V1, thereby ensuring that semantic security remains unaltered.
Consequently, the focus is directed towards proving that the \ac{PKE} of both versions of \ac{WKyber} are \ac{IND-CPA} secure.
The security of Kyber's mathematical foundation derives from the fact that Kyber \ac{PKE} generates keys and ciphertexts that are instances of the \ac{LWE} problem, meaning that its security is based on the computational difficulty of solving these instances.
Kyber \ac{PKE}'s security reduction can be also be considered for the two versions of \ac{WKyber} \ac{PKE}, but only if the hardness of the \ac{LWE} problem is equivalent for the \ac{AWGN} channel error distribution.

The \ac{LWE} and \ac{MLWE} problems are stated with parameters $q$, $n$, $k$ and $(\eta_1$, $\eta_2)$, and the additional hidden parameter $m$.
The $\eta_i$ parameters denote the range of the error distributions used in the cryptosystem.
In the \ac{ML-KEM} documentation \cite{FIPS:2024:203}, the error distribution is defined as a discrete binomial  distribution centered at the origin, with values in the interval $[-\eta_i, \eta_i]$.
However, the general formulation of the \ac{LWE} problem models, the error distribution as a centered Gaussian distribution.
The first version of Kyber submitted to \ac{NIST} followed this convention, but in the second round of the \ac{NIST} \ac{PQC} call, the authors proposed to use a centered binomial distribution to sample the \ac{LWE} errors, as its implementation is much easier and does not harm the security of the cryptosystem.
In the third round submission of Kyber \cite{NIST:3R:kyber}, it is stated that the execution of the best attacks against the cryptosystem does not depend on the nature of the error distribution, but on its standard deviation.
Before Kyber, another lattice-based cryptosystem called NewHope proposed the change of error distribution, from a Gaussian to a binomial in \cite{NewHope-USENIX}. The literature further presents examples of works changing the model of the error distribution, like FrodoKEM \cite{NIST:3R:Frodo} that defines a specific error distribution for this cryptosystem.
As stated in Lemma 5.5 of the FrodoKEM documentation, the alteration in error distribution requires an additional security reduction.

To solve the \ac{LWE} problem, two attacks are considered to be viable strategies 1) the dual and 2) the primal attack, where in the case of Kyber, the authors considered two estimation strategies.
First, the Core-\ac{SVP} \cite{NewHope-USENIX} primal attack strategy was included as a cost measurement in Kyber's security analysis, which consists in only taking into account the cost of a single call to an \ac{SVP} oracle in a fixed dimension.
Therefore, it is seen as a pessimistic estimation.

\begin{figure}[h]
\centering
\includegraphics[width=\columnwidth]{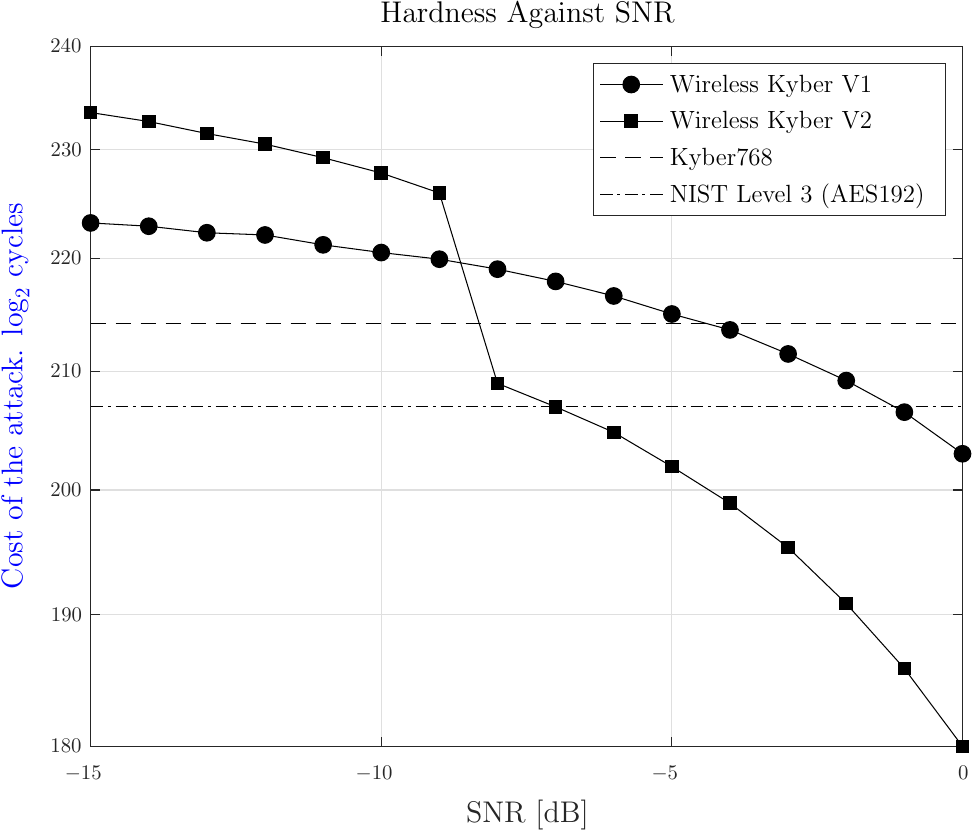}
\caption{ Cost of the dual attack against \ac{LWE} samples of \ac{WKyber} \ac{PKE} V1 and V2 in comparison to standard Kyber and NIST level 3 security bound (\ac{AES}192)}
\label{fig:Sec-WKyber}
\vspace{1ex}
\includegraphics[width=\columnwidth]{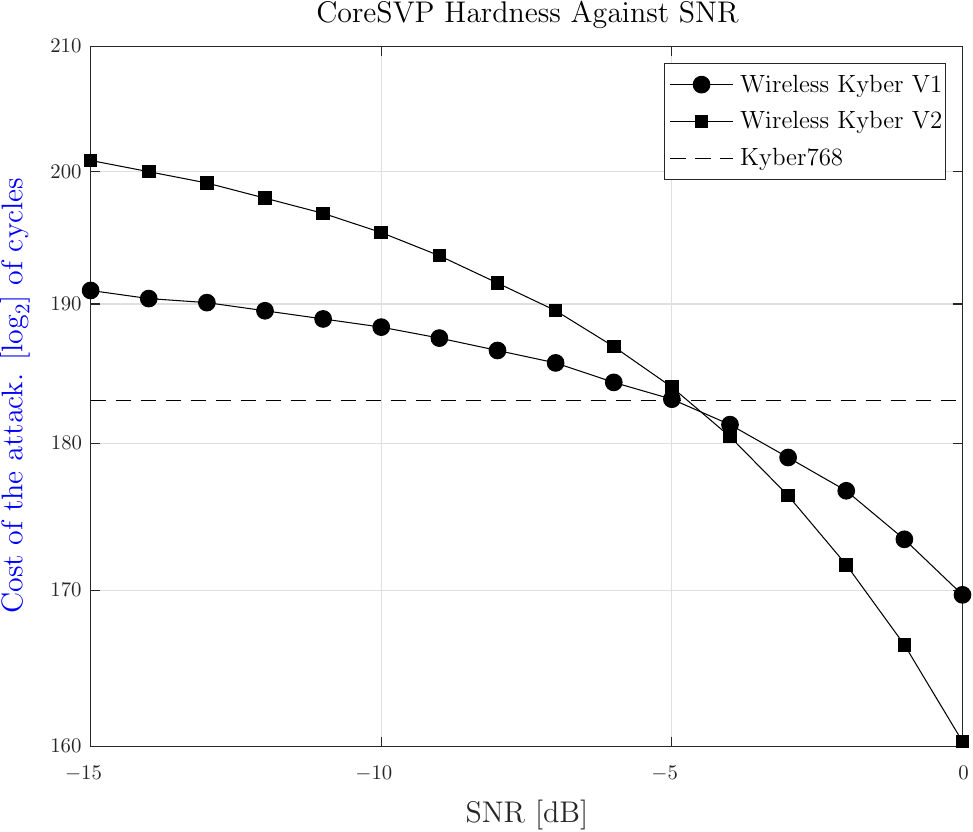}
\caption{ Cost of using the primal CoreSVP strategy against \ac{LWE} samples of \ac{WKyber} \ac{PKE} V1 and V2 in comparison to standard Kyber}
\label{fig:Sec-WKyber-2}
\end{figure}

The dual attack \cite{Albrecht:2015:Estimator, Albrecht:2022:QuantumAD} is  the second main strategy to solve a \ac{LWE} instance, in particular the \ac{LWE} decision problem.
The cost of the dual attack gives an estimation that fits better the comparison with the \ac{NIST} security levels.
In contrast, the estimations for the original Kyber with the Core\ac{SVP} methodology do not reach the corresponding target level of security on any parameter set.
Both Kyber and \ac{ML-KEM} estimate the security level of each parameter set as the hardness associated with such parameters against the dual attack, i.e. cost of running the attack.

To obtain an equivalent security evaluation for \ac{WKyber}, the open source lattice estimator presented in \cite{Albrecht:2024:LE, Albrecht:2015:Estimator} was used.
It provides various estimations of the complexity of solving a designated \ac{LWE} instance, based on diverse strategies, criteria, and publications.
This tool facilitates the selection of a \ac{LWE} parameter set, with the option to define distinct distributions of error.
In particular, the cost of the dual or primal attacks can be estimated for any given \ac{LWE} parameter set.
Additionally, the calculus of the cost of other attacks, based on more recent publications, is included in the estimator.
The hybrid attack, proposed in \cite{Espitau:2020:OnAD}, is relevant due to its better performance in some cases, hence it is included in the security analysis.

The \ac{AWGN} channel distribution considered in \ac{WKyber} was introduced in the lattice estimator \cite{Albrecht:2015:Estimator} as a \ac{LWE} error distribution that generates $\bm{e}$, $\bm{e}'$ and $e''$.
The cost of the dual, primal and hybrid attack was calculated using the estimator, creating \ac{LWE} parameters corresponding to \ac{WKyber} \ac{PKE} V1, \ac{WKyber} \ac{PKE} V2 and Kyber \ac{PKE}.
Tables~\ref{tb:Security-WK1} and \ref{tb:Security-WK2} show the results of the estimator, including between brackets, the cost of the attacks against Kyber presented in the reference documentation.

To analyze the effect of the change of error distribution, Table~\ref{tb:Security-WK1} compares the computational cost of solving the \ac{LWE} problem using the attacks previously presented against the instances of Kyber \ac{PKE} and \ac{WKyber} \ac{PKE} V1, the latter transmitted with a \ac{SNR} of $-10$~dB.
Although this particular \ac{SNR} was chosen for analysis, it is important to note that other values could also be selected.
This can be observed in Figures~\ref{fig:Sec-WKyber} and \ref{fig:Sec-WKyber-2}, which illustrate that the change in the error distribution does not compromise the hardness of \ac{LWE} instances up to \ac{SNR} of $-5$~dB.
The cost of the attacks is evaluated in terms of CPU cycles; also, some specific algorithms have also a exponential cost of memory.

When the error distribution of the \ac{PQ}-cryptosystems Kyber or NewHope was modified \cite{NIST:3R:kyber, NewHope-USENIX}, it was argued that the security was unaffected as long as the standard deviation remained the same.
This argument is reinforced in the estimator presented in \cite{Albrecht:2015:Estimator}, since only the standard deviation is used in the calculus of the cost of solving LWE.
Therefore, if the distribution is changed to a different one with higher standard deviation, the cost of solving the problem also rises.
The binomial distribution presented in the original parameter sets has a standard deviation of $1$.
The standard deviation of the channel error distribution is determined by Eq.~(\ref{eq:pe_coeff}), depending upon the \ac{SNR}.
Consequently, the estimations of the cost of solving \ac{WKyber} instances vary according to the \ac{SNR} utilized to transmit the two least significant bits.
A lower standard deviation corresponds to a lower cost of attacks, and conversely, a higher \ac{SNR} is associated with a lower security estimation.
The \ac{SNR} considered for the results presented in Figures \ref{fig:Sec-WKyber} and \ref{fig:Sec-WKyber-2} ranges from $-15$~dB to $0$~dB.

These results demonstrate that under optimal SNR conditions the security of \ac{WKyber} is equivalent to that of standard Kyber.
The cost of the dual attack as a function of the \ac{SNR} is presented in Figure~\ref{fig:Sec-WKyber}.
\ac{WKyber} \ac{PKE} V1 and V2 achieve a higher degree of security than Kyber for an \ac{SNR} in the ranges $[-15$~dB, $-5$~dB$]$, and $[-15$~dB, $-9$~dB$]$ respectively.
Furthermore, a comparison between the hardness of \ac{WKyber} \ac{LWE} instances and the computational cost of solving \ac{AES}192 is presented, revealing a wider range of secure \ac{SNR} values.

\begin{table}[H]
\begin{center}
\caption{Hardness (log$_2$ of cycles) of the instances presented by \Ac{SotA} Kyber and Proposed \ac{WKyber} \ac{PKE} V1 at -10dB\tablefootnote{Additional \ac{SNR} values for Hardness and CoreSVP Hardness can be seen in Figures \ref{fig:Sec-WKyber} and \ref{fig:Sec-WKyber-2}}}
\label{tb:Security-WK1}
\resizebox{7.6cm}{!}{
\begin{tabular}{|@{\,}c@{\,}|@{\,}c@{\,}|@{\,}c@{\,}|} \hline
 & Kyber512 & $ \text{\ac{WKyber} V1}\ k = 2$ \\ \hline\hline
CoreSVP classical  & $115 (118)$ & $119.4$ \\ \hline
Dual attack & $146.2 (151.5)$  & $150.9$ \\ \hline
Dual hybrid attack & $135.3$ & $140$ \\ \hline \hline
 & Kyber768 & $ \text{\ac{WKyber} V1}\ k = 3$ \\ \hline\hline
CoreSVP classical  & $182.2$ & $188$ \\ \hline
Dual attack & $214.2 (215.1)$  & $220.5$ \\ \hline
Dual hybrid attack & $196.1$ & $190.3$ \\ \hline \hline
 & Kyber1024 & $ \text{\ac{WKyber} V1}\ k = 4$ \\ \hline\hline
CoreSVP classical  & $255.2 (256)$ & $263.1$ \\ \hline
Dual attack & $288.6 (287.3)$  & $296.9$ \\ \hline
Dual hybrid attack & $262.4$ & $268.5$ \\ \hline
\end{tabular}
}
\vspace{1ex}
\caption{Hardness (log$_2$ of cycles) of the instances presented by \Ac{SotA} Kyber and Proposed \ac{WKyber} \ac{PKE} V2 at -10dB$^\text{\normalfont 3}$}
\label{tb:Security-WK2}
\vspace{-2ex}
\resizebox{7.6cm}{!} {
\begin{tabular}{|@{\,}c@{\,}|@{\,}c@{\,}|@{\,}c@{\,}|} \hline
& Kyber512 & $\text{\ac{WKyber} V2}\ k = 2$ \\ \hline\hline
CoreSVP classical  & $115 (118)$ & $120$ \\ \hline
Dual attack & $146.2 (151.5)$  & $151.5$ \\ \hline
Dual hybrid attack & $135.3$ & $140.8$ \\ \hline \hline
& Kyber768 & $\text{\ac{WKyber} V2}\ k = 2$ \\ \hline\hline
CoreSVP classical  & $182.2$ & $195.3$ \\ \hline
Dual attack & $214.2 (215.1)$  & $227.8$ \\ \hline
Dual hybrid attack & $196.1$ & $208.9$ \\ \hline \hline
& Kyber1024 & $\text{\ac{WKyber} V2}\ k = 2$ \\ \hline\hline
CoreSVP classical  & $255.2 (256)$ & $272.7$ \\ \hline
Dual attack & $288.6 (287.3)$  & $ 306.5$ \\ \hline
Dual hybrid attack & $262.4$ & $279.3$ \\ \hline
\end{tabular}}
\end{center}
\vspace{-3ex}
\end{table}

Figure~\ref{fig:Sec-WKyber-2} shows the cost of the primal attack against WKyber \ac{PKE} and the standard Kyber, assuming the Core\ac{SVP} strategy. As can be appreciated, the results for \ac{SNR}  lower than $-5$~dB are, again, better than the respective results of Kyber, thereby ensuring a consistent level of security.

The analysis indicates that the standard deviation significantly impacts the performance of cryptanalysis against \ac{LWE}. As previously explained and further elaborated in the Kyber submission, the performance of the cryptanalysis with respect to the error distribution is primarily influenced by the standard deviation. The binomial distribution employed in Kyber768 has a standard deviation of 1, and for \ac{SNR} $\geq -5$~dB, the standard deviation of the error distribution of the channel exceeds 1.

\vspace{-1ex}
\subsection{Error probability}

Kyber/ML-KEM are probabilistic cryptosystems,  a characteristic shared by the majority of LWE-based cryptosystems, this implies an inherent probability of error in the exchange of information and data, which has a significant impact on the final error probability  of the cryptosystem. Therefore, this section is dedicated on analyzing the error probability of the \ac{WKyber} cryptosystem.

\begin{table}[H]
\begin{center}
\caption{Error probability of Kyber and \ac{WKyber} \ac{PKE} V1/V2 ($\text{SNR}\!=\!-10$dB)}
\label{tb:Error-Prob}
\vspace{-1ex}
{\small
\begin{tabular}{|c|c|} \hline
& Error probability ($\log_2$)\\ \hline
Kyber512 & -139 \\
WKyber V1 $k=2$ & -219.1 \\
WKyber V2 $k=2$ & -198.7 \\ \hline
Kyber768 & -164 \\
WKyber V1 $k=3$ & -227.2 \\
WKyber V2 $k=3$ & -138.5 \\ \hline
Kyber1024 & -175\\
WKyber V1 $k=4$ & -174 \\
WKyber V2 $k=4$ & -105.2 \\ \hline
\end{tabular}
}
\end{center}
\vspace{-4ex}
\end{table}

The designers of Kyber present a Python script to calculate the error probability of the cryptosystem.
This script was adapted to include the change in error distribution and the absence of compression.
The change in the error distribution increases the probability of error; however, not applying the compression significantly mitigates this effect.
As expected, the error probability of \ac{WKyber} is higher than the one from the original cryptosystem. This is due to the fact that the range of the binomial distribution used in Kyber is $[-2,2]$, while the channel error distribution range varies with the \ac{SNR}.
The security analysis in Subsection~\ref{ss:security} shows that if the error distribution has a wider range, the complexity of the LWE instances is enhanced.
In the case of WKyber, the security in terms of cost of attacks is higher than the same of Kyber, but also the error probability is higher.
Not using the $\text{Compress}_q$ function on ciphertexts means a loss in performance, given the necessity to transmit longer bit strings.
However, from a security standpoint the absence of compression results in a reduction in the error probability.

The error probability of Kyber derives from the following expression
\vspace{-1ex}
\begin{equation}
\text{Decompress}_q(m,1) + \bm{e}^\mathrm{T} \bm{s}' - \bm{s}^\mathrm{T} \bm{e}' + e'' + e_u + e_v,
\vspace{-1ex}
\end{equation}
where the coefficients of $\bm{e}$, $\bm{e}'$, $\bm{s}$, $\bm{s}'$, and $e''$ are sampled using the error distribution of the cryptosystem, and $e_u$ and $e_v$ represent the error introduced during the compression. In the case of \ac{WKyber}, as discussed in section~\ref{ss:pack-and-cod}, it is important to note that the compression is not applied, hence this error is not considered. Since the error of compression is not considered in the security analysis of the original submission of Kyber \cite{NIST:3R:kyber}, the omission of this function does not affect the security of \ac{WKyber}. Table \ref{tb:Error-Prob} shows the probability of error in a key exchange for both, the original Kyber and the proposed WKyber versions. It can be appreciated that the version 1 of \ac{WKyber} exhibits a lower error probability than the original scheme. On the other hand, the version 2 of \ac{WKyber} presents a higher error probability than Kyber, however this can be mitigated using a higher SNR still in the secure range between $-10$~dB and $-5$~dB.

\begin{figure}[H]
\centering
\includegraphics[width=\columnwidth]{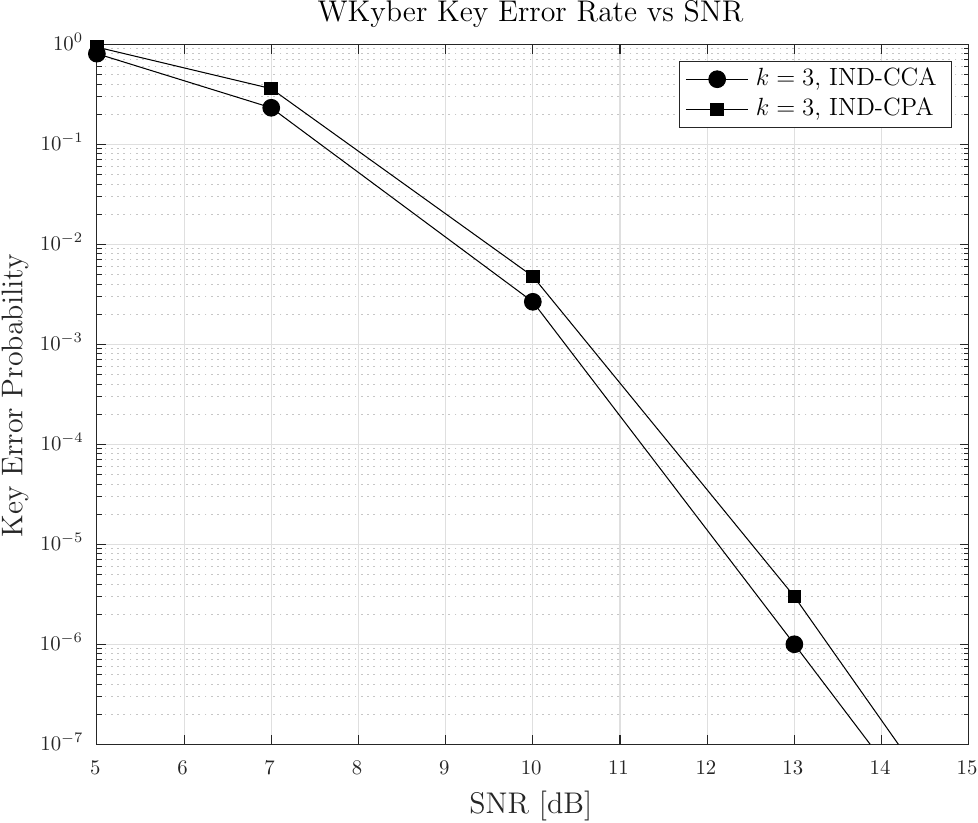}
\caption{\ac{KER} for the \ac{WKyber} 768 V1/V2 system under various SNRs and values of $k$}
\label{fig:KER}
\vspace{-2ex}
\end{figure}

Despite the inclusion of the \ac{BCH} code for error correction, there is still a chance that will not be able to correct all errors.
As presented in \ref{ss:pack-and-cod}, the first 10 bits of each polynomial coefficient are coded before their transmission, adding redundant bits which add the capacity to correct up to $5$ errors.
 The probability of a codeword not being correctly decoded was defined in Eq.~\eqref{eq:Pce}, while Figure~\ref{fig:KER} represents the errors of the key exchange vs. the \ac{SNR}.

 The schemes presented in Tables~\ref{tb:WKyber-V1} and \ref{tb:WKyber-V2} define two \ac{SNR} values on each communication, one for the ten most significant bits of each coefficient, which controls the decoding error probability, and one for the two least significant bits, which control the hardness of the system.
As demonstrated in Figure~\ref{fig:KER}, utilizing higher \ac{SNR} results in a reduced error probability.
It is notable that, for \ac{SNR}~$= 15$~dB, the error function falls in Figure~\ref{fig:KER} under the numerical underflow.
In summary, managing the power profile utilized during a \ac{WKyber} is imperative avoid errors and maintain security, as the \ac{SNR} achieved for the \ac{BCH} words must be greater than $10$~dB, and the one for the last two bits of each coefficient are transmitted to be below $-5$~dB to avoid security losses, and keep negligible the error probability of the scheme.

\vspace{-1ex}
\subsection{A Note on Implementation Requirements}

In this subsection the feasibility of the proposed \ac{WKyber} schemes is analyzed.
The first requirement to implement \ac{WKyber} is the adaptation of the underlying mathematical functions, like \ac{NTT} multiplication and other lattice arithmetics, to a physical/hardware implementation; ideally, an implementation contained in a hardware security module.
This is, however, a well known reality for the adoption of post-quantum cryptography, specially in IoT. As hardware implementations of cryptographic primitives generally perform better, those are currently demanded for some implementation of Kyber.
In this regard, therefore, it can be said that \ac{WKyber} demands no additional cryptographic requirements compared to currently used methods.

Next, in order to address the overhead associated with implementing \ac{WKyber} in comparison to standard Kyber, one can look into the efficiency of both approaches with respect to requirements in terms of transmitted information.
Considering an implementation over Bluetooth 5.0 as an example, the efficiency of the \ac{WKyber} scheme can be estimated at $\frac{12}{16 + 2}\approx 67\%$, which follows from the encoding scheme employed, namely, BCH coding of 10 bit codewords into 15 bits, rounded up to 16 due to the 4QAM modulation, plus the lower 2-bit word.
In comparison, it is known \cite{AfanehBLE2023, XuBLEComL2024, WoolleyBluetooth2025} that under 1 MBit/s and 2 Mbit/s raw transmission, \ac{BLE} achieves efficiencies of 23\%--48\% bits, and 19\%--68\% bits, respectively.
In other words, the spectral efficiency of \ac{WKyber} can be said to be comparable to the highest achievable by the most recent and advanced version of \ac{BLE}, which is one of the key wireless technologies for \ac{IoT} applications.

The standard Kyber scheme requires the implementation of an auxiliary error function in order to generate the errors added to the messages and keys.
In contrast, under \ac{WKyber} V1, the call to the binomial distribution for the error of the ciphertext is not required, while under \ac{WKyber} V2 the equivalent call is also not needed during the execution of the key generation algorithm, which implicates that \ac{WKyber} is more efficient and can run faster than standard Kyber.

Finally, it should be noted that issues such as added latency by using a \ac{PQ} scheme are also mitigated due to the efficiency of communication mentioned earlier, meaning that standard Kyber requires error-free channel, while \ac{WKyber} does not.
As for the latency associated with key establishment, this is inevitable when adopting \ac{PQ} schemes \cite{FIPS:2024:203}, as it is already well known that keys, signatures and the messages needed to exchange them are longer than those in pre-quantum cryptography \cite{FIPS:2018:800, FIPS:2023:186-5}.
In other words, by limiting the overhead at the transport layer, the overall impact is lowered, and latency gains/losses due to computational complexity of the \ac{WKyber} algorithm are not relevant, since the highest complexity operations are the generation of the public key matrix $A$ and its multiplication with the generated secret.

\section{Conclusions}
\label{S:Conclusion}

The field of crypto-security is currently at a compromised and convulsive moment.
Various authorities and standardization bodies are calling for a transition to post-quantum cryptography and the adoption of hybrid systems.
Following this trend, we propose a combination of the first standard adopted by NIST, CRYSTALS-Kyber, and the use of the physical security layer.
Public key encryption is among the subjects most affected by the quantum threat and algorithms such as ElGamal, RSA and those based on elliptic curves will no longer be secure.
The proposal of Wireless Kyber can be seen as a step in the post-quantum transition with the possibility of replacing those algorithms in wireless IoT devices.
A viable adaptation makes use of BCH codes in the error correction section,  and of signal strength manipulation to introduce an error in each coefficient, very similar to that of the Kyber cryptosystem.
Finally, the security and error probability of \ac{WKyber} have been analyzed to assess the feasibility of the cryptosystem, with the conclusion that if the SNR difference is sufficiently high, a secure exchange, analogous to the original Kyber key exchange, can be maintained.
The implementation refinement and analysis of other vulnerabilities are left as open work.




\begin{thebibliography}{10}
\providecommand{\url}[1]{#1}
\csname url@samestyle\endcsname
\providecommand{\newblock}{\relax}
\providecommand{\bibinfo}[2]{#2}
\providecommand{\BIBentrySTDinterwordspacing}{\spaceskip=0pt\relax}
\providecommand{\BIBentryALTinterwordstretchfactor}{4}
\providecommand{\BIBentryALTinterwordspacing}{\spaceskip=\fontdimen2\font plus
\BIBentryALTinterwordstretchfactor\fontdimen3\font minus
  \fontdimen4\font\relax}
\providecommand{\BIBforeignlanguage}[2]{{%
\expandafter\ifx\csname l@#1\endcsname\relax
\typeout{** WARNING: IEEEtran.bst: No hyphenation pattern has been}%
\typeout{** loaded for the language `#1'. Using the pattern for}%
\typeout{** the default language instead.}%
\else
\language=\csname l@#1\endcsname
\fi
#2}}
\providecommand{\BIBdecl}{\relax}
\BIBdecl

\bibitem{ericsson19}
Ericsson, ``Ericsson mobility report,'' Ericsson, Tech. Rep., 2019,
  \url{https://www.ericsson.com/4acd7e/assets/local/reports-papers/mobility-report/documents/2019/emr-november-2019.pdf}.

\bibitem{fang18}
D.~Fang, Y.~Qian, and R.~Q. Hu, ``Security for {5G} mobile wireless networks,''
  \emph{{IEEE} Access}, vol.~6, pp. 4850--4874, 2018,
  \url{https://doi.org/10.1109/ACCESS.2017.2779146}.

\bibitem{ahmad19}
I.~Ahmad, S.~Shahabuddin, T.~Kumar, J.~Okwuibe, A.~Gurtov, and M.~Ylianttila,
  ``Security for {5G} and beyond,'' \emph{{IEEE} Communications Surveys \&
  Tutorials}, vol.~21, no.~4, pp. 3682--3722, 2019,
  \url{https://doi.org/10.1109/COMST.2019.2916180}.

\bibitem{deutsch1985quantum}
D.~Deutsch and R.~Penrose, ``Quantum theory, the {C}hurch--{T}uring principle
  and the universal quantum computer,'' \emph{Proc. Royal Society London. A.
  Mathematical and Physical Sciences}, vol. 400, no. 1818, pp. 97--117, 1985,
  \url{https://doi.org/10.1098/rspa.1985.0070}.

\bibitem{cross2019validating}
A.~W. Cross, L.~S. Bishop, S.~Sheldon, P.~D. Nation, and J.~M. Gambetta,
  ``Validating quantum computers using randomized model circuits,''
  \emph{Physical Review A}, vol. 100, no.~3, p. 032328, 2019,
  \url{https://doi.org/10.1103/PhysRevA.100.032328}.

\bibitem{cumming2022using}
R.~Cumming and T.~Thomas, ``Using a quantum computer to solve a real-world
  problem -- what can be achieved today?'' \emph{arXiv}, no. arXiv:2211.13080,
  2022, \url{https://arxiv.org/abs/2211.13080}.

\bibitem{ichikawa2023comprehensive}
T.~Ichikawa, H.~Hakoshima, K.~Inui, K.~Ito, R.~Matsuda, K.~Mitarai,
  K.~Miyamoto, W.~Mizukami, K.~Mizuta, T.~Mori, Y.~Nakano, A.~Nakayama, K.~N.
  Okada, T.~Sugimoto, S.~Takahira, N.~Takemori, S.~Tsukano, H.~Ueda,
  R.~Watanabe, Y.~Yoshida, and K.~Fujii, ``A comprehensive survey on quantum
  computer usage: How many qubits are employed for what purposes?''
  \emph{arXiv}, no. arXiv:2307.16130, 2023,
  \url{https://arxiv.org/abs/2307.16130}.

\bibitem{Shor:1997:PTA}
P.~W. Shor, ``Polynomial-time algorithms for prime factorization and discrete
  logarithms on a quantum computer,'' \emph{SIAM Review}, vol.~41, no.~2, pp.
  303--332, 1999, \url{https://doi.org/10.1137/S0036144598347011}.

\bibitem{FIPS:2024:203}
{NIST}, \emph{Module-Lattice-Based Key-Encapsulation Mechanism Standard},
  National Institute of Standard and Technology, FIPS 203, 2024,
  \url{https://doi.org/10.6028/NIST.FIPS.203}.

\bibitem{NIST:3R:kyber}
R.~Avanzi, J.~Bos, E.~Kiltz, T.~Lepoint, V.~Lyubashevsky, J.~M. Schanck,
  P.~Schwabe, G.~Seiler, and D.~Stehl\'e, ``{CRYSTALS-Kyber},'' Online
  publication, 2020, \url{https://pq-crystals.org}.

\bibitem{MaringerIFS2023}
G.~Maringer, S.~Puchinger, and A.~Wachter-Zeh, ``Information- and
  coding-theoretic analysis of the {RLWE/MLWE} channel,'' \emph{IEEE
  Transactions on Information Forensics and Security}, vol.~18, pp. 549--564,
  2023, \url{https://doi.org/10.1109/TIFS.2022.3226907}.

\bibitem{WangISPA2015}
M.~Wang and Z.~Yan, ``Security in {D2D} communications: A review,'' in
  \emph{2015 IEEE Trustcom/BigDataSE/ISPA}, vol.~1, 2015, pp. 1199--1204,
  \url{https://doi.org/10.1109/Trustcom.2015.505}.

\bibitem{AndreaISCC2015}
I.~Andrea, C.~Chrysostomou, and G.~Hadjichristofi, ``Internet of things:
  Security vulnerabilities and challenges,'' in \emph{2015 IEEE Symposium on
  Computers and Communication (ISCC)}, 2015, pp. 180--187,
  \url{https://doi.org/10.1109/ISCC.2015.7405513}.

\bibitem{JungITJ2022}
J.~Jung, B.~Kim, J.~Cho, and B.~Lee, ``A secure platform model based on {ARM}
  platform security architecture for {IoT} devices,'' \emph{IEEE Internet of
  Things Journal}, vol.~9, no.~7, pp. 5548--5560, 2022,
  \url{https://doi.org/10.1109/JIOT.2021.3109299}.

\bibitem{HuangIFS2024}
J.~Huang, H.~Zhao, J.~Zhang, W.~Dai, L.~Zhou, R.~C.~C. Cheung, C.~K. Ko\c{c},
  and D.~Chen, ``Yet another improvement of {Plantard} arithmetic for faster
  {Kyber} on low-end 32-bit {IoT} devices,'' \emph{IEEE Transactions on
  Information Forensics and Security}, vol.~19, pp. 3800--3813, 2024,
  \url{https://doi.org/10.1109/TIFS.2024.3371369}.

\bibitem{porambage21}
P.~Porambage, G.~G{\"u}r, D.~P. Moya~Osorio, M.~Livanage, and M.~Ylianttila,
  ``{6G} security challenges and potential solutions,'' in \emph{2021 Joint
  European Conference on Networks and Communications \& 6G Summit (EuCNC/{6G}
  Summit)}, 2021, pp. 622--627,
  \url{https://doi.org/10.1109/EuCNC/6GSummit51104.2021.9482609}.

\bibitem{MucchiOJCS2021}
L.~Mucchi, S.~Jayousi, S.~Caputo, E.~Panayirci, S.~Shahabuddin, J.~Bechtold,
  I.~Morales, R.-A. Stoica, G.~Abreu, and H.~Haas, ``Physical-layer security in
  {6G} networks,'' \emph{IEEE Open Journal of the Communications Society},
  vol.~2, pp. 1901--1914, 2021,
  \url{https://doi.org/10.1109/OJCOMS.2021.3103735}.

\bibitem{AngueiraCST2022}
P.~Angueira, I.~Val, J.~Montalb{\'a}n, {\'O}.~Seijo, E.~Iradier, P.~S.
  Fontaneda, L.~Fanari, and A.~Arriola, ``A survey of physical layer techniques
  for secure wireless communications in industry,'' \emph{IEEE Communications
  Surveys \& Tutorials}, vol.~24, no.~2, pp. 810--838, 2022,
  \url{https://doi.org/10.1109/COMST.2022.3148857}.

\bibitem{katsuki2022noncoherent}
Y.~Katsuki, G.~T. F.~d. Abreu, K.~Ishibashi, and N.~Ishikawa, ``Noncoherent
  massive {MIMO} with embedded one-way function physical layer security,''
  \emph{IEEE Transactions on Information Forensics and Security}, vol.~18, pp.
  3158--3170, 2023, \url{https://doi.org/10.1109/TIFS.2023.3277255}.

\bibitem{wyner1975wire}
A.~D. Wyner, ``The wire-tap channel,'' \emph{Bell system Technical Journal},
  vol.~54, no.~8, pp. 1355--1387, 1975,
  \url{https://doi.org/10.1109/IISA.2014.6878717}.

\bibitem{senlin24}
S.~Liu, T.~Gao, Y.~Liu, and X.~Lu, ``Physical-layer public key encryption
  through massive {MIMO},'' in \emph{Proc. 19th ACM Asia Conference on Computer
  and Communications Security}.\hskip 1em plus 0.5em minus 0.4em\relax New
  York, NY, USA: Association for Computing Machinery, 2024,
  \url{https://doi.org/10.1145/3634737.3656284}.

\bibitem{ABDALLAH24}
W.~Abdallah, ``A physical layer security scheme for {6G} wireless networks
  using post-quantum cryptography,'' \emph{Computer Communications}, vol. 218,
  pp. 176--187, 2024, \url{https://doi.org/10.1016/j.comcom.2024.02.019}.

\bibitem{TosunIFS2024}
T.~Tosun and E.~Savas, ``Zero-value filtering for accelerating non-profiled
  side-channel attack on incomplete {NTT}-based implementations of
  lattice-based cryptography,'' \emph{IEEE Transactions on Information
  Forensics and Security}, vol.~19, pp. 3353--3365, 2024,
  \url{https://doi.org/10.1109/TIFS.2024.3359890}.

\bibitem{NosouhiIFS2023}
M.~R. Nosouhi, S.~W. Shah, L.~Pan, Y.~Zolotavkin, A.~Nanda, P.~Gauravaram, and
  R.~Doss, ``Weak-key analysis for {BIKE} post-quantum key encapsulation
  mechanism,'' \emph{IEEE Transactions on Information Forensics and Security},
  vol.~18, pp. 2160--2174, 2023,
  \url{https://doi.org/10.1109/TIFS.2023.3264153}.

\bibitem{Peikert:2009:PKC}
C.~Peikert, ``Public-key cryptosystems from the worst-case shortest vector
  problem: Extended abstract,'' in \emph{Proc. Forty-First Annual ACM Symposium
  on Theory of Computing}, 2009, pp. 333--342,
  \url{https://doi.org/10.1145/1536414.1536461}.

\bibitem{RaviIFS2022}
P.~Ravi, S.~Bhasin, S.~S. Roy, and A.~Chattopadhyay, ``On exploiting message
  leakage in (few) {NIST PQC} candidates for practical message recovery
  attacks,'' \emph{IEEE Transactions on Information Forensics and Security},
  vol.~17, pp. 684--699, 2022, \url{https://doi.org/10.1109/TIFS.2021.3139268}.

\bibitem{Jiang:2018:FO}
H.~Jiang, Z.~Zhang, L.~Chen, H.~Wang, and Z.~Ma, ``{IND-CCA}-secure key
  encapsulation mechanism in the quantum random oracle model, revisited,'' in
  \emph{Annual International Cryptology Conference, Lecture Notes Comput.
  Sci.,}, vol. 10993, 2018, pp. 96--125,
  \url{https://doi.org/10.1007/978-3-319-96878-0_4}.

\bibitem{tse}
D.~Tse and P.~Viswanath, \emph{Fundamentals of wireless Communication}.\hskip
  1em plus 0.5em minus 0.4em\relax Cambridge University Press, 2008.

\bibitem{DumanBookCh2007}
T.~M. Duman and A.~Ghrayeb, \emph{Fading Channels and Diversity Techniques},
  2007, pp. 7--42, \url{https://doi.org/10.1002/9780470724347.ch2}.

\bibitem{BIGLIERI20001135}
E.~Biglieri, G.~Caire, and G.~Taricco, ``Coding for the fading channel: a
  survey,'' \emph{Signal Processing}, vol.~80, no.~7, pp. 1135--1148, 2000,
  \url{https://doi.org/10.1016/S0165-1684(00)00027-X}.

\bibitem{Proakis:2007:DCom}
J.~G. Proakis and M.~Salehi, \emph{Digital Communications}.\hskip 1em plus
  0.5em minus 0.4em\relax 5th Edition, McGraw Hill, 2007,
  \url{https://daskalakispiros.com/files/Ebooks/digital-communication-proakis-salehi-5th-edition.pdf}.

\bibitem{bc60}
R.~Bose and D.~Ray-Chaudhuri, ``On a class of error correcting binary group
  codes,'' \emph{Information and Control}, vol.~3, no.~1, pp. 68--79, 1960,
  \url{https://doi.org/10.1016/S0019-9958(60)90287-4}.

\bibitem{h59}
A.~Hocquenghem, ``Codes correcteurs d'erreurs,'' \emph{Chiffres}, 1959.

\bibitem{lin_error}
S.~Lin and D.~J. Costello, \emph{Error Control Coding}.\hskip 1em plus 0.5em minus
  0.4em\relax Prentice Hall, 2004.

\bibitem{Hofheinz:2017:MAF}
D.~Hofheinz, K.~H{\"{o}}velmanns, and E.~Kiltz, ``A modular analysis of the
  {Fujisaki}-{Okamoto} transformation,'' in \emph{Proc. 15th International
  Conference Theory of Cryptography {TCC}'2017, Lecture Notes Comput. Sci.},
  vol. 10677, 2017, pp. 341--371,
  \url{https://doi.org/10.1007/978-3-319-70500-2_12}.

\bibitem{Nist:2023:Draft1}
{NIST}, \emph{Module-Lattice-based Key Encapsulation Mechanism Standard},
  National Institute of Standards and Tecnology, FIPS 203 ipd., 2023,
  \url{https://doi.org/10.6028/NIST.FIPS.203.ipd}.

\bibitem{NewHope-USENIX}
E.~Alkim, L.~Ducas, T.~P\"oppelmann, and P.~Schwabe, ``Post-quantum key
  exchange - {A} new hope,'' in \emph{Proc. 25th USENIX Security Symposium},
  2016, pp. 327--343,
  \url{https://www.usenix.org/system/files/conference/usenixsecurity16/sec16_paper_alkim.pdf}.

\bibitem{NIST:3R:Frodo}
E.~Alkim, J.~W. Bos, L.~Ducas, P.~Longa, I.~Mironov, M.~Naehrig, V.~Nikolaenko,
  C.~Peikert, A.~Raghunathan, and D.~Stebila, ``{FrodoKEM} learning with errors
  key encapsulation ({Round} 3 {Submission}),'' Online publication, 2021,
  \url{https://frodokem.org/#spec}.

\bibitem{Albrecht:2015:Estimator}
M.~R. Albrecht, R.~Player, and S.~Scott, ``On the concrete hardness of learning
  with errors,'' \emph{J. Math. Cryptol.}, vol.~9, no.~3, pp. 169--203, 2015,
  \url{https://doi:10.1515/jmc-2015-0016}.

\bibitem{Albrecht:2022:QuantumAD}
M.~R. Albrecht and Y.~Shen, ``Quantum augmented dual attack,'' \emph{Cryptology
  ePrint Archive, Report 2002/656}, 2022,
  \url{https://eprint.iacr.org/2022/656}.

\bibitem{Albrecht:2024:LE}
M.~R. Albrecht, ``Lattice-estimator,'' On-line publication, 2024,
  \url{https://github.com/malb/lattice-estimator/tree/main}.

\bibitem{Espitau:2020:OnAD}
T.~Espitau, A.~Joux, and N.~Kharchenko, ``On a dual/hybrid approach to small
  secret {LWE} - a dual/enumeration technique for learning with errors and
  application to security estimates of {FHE} schemes,'' in \emph{International
  Conference on Cryptology in India}, 2020,
  \url{https://api.semanticscholar.org/CorpusID:219332962}.

\bibitem{AfanehBLE2023}
M.~Afaneh, ``Bluetooth 5 speed: How to achieve maximum throughput for your
  {BLE} application,'' 2023,
  \url{https://novelbits.io/bluetooth-5-speed-maximum-throughput}.

\bibitem{XuBLEComL2024}
H.~Xu, Z.~Yan, B.~Li, and M.~Yang, ``Modeling and analysis of the performance
  for {Bluetooth} low energy,'' \emph{IEEE Communications Letters}, vol.~28,
  no.~3, pp. 732--736, 2024, \url{https://doi.org/10.1109/LCOMM.2024.3352545}.

\bibitem{WoolleyBluetooth2025}
M.~Woolley, ``Bluetooth core 5.0 feature enhancements,'' Bluetooth, Tech. Rep.,
  2025,
  \url{https://www.bluetooth.com/bluetooth-resources/bluetooth-core-5-0-go-faster-go-further/}.

\bibitem{FIPS:2018:800}
{NIST}, \emph{Recommendation for Pair-Wise Key-Establishment Schemes Using
  Discrete Logarithm Cryptography}, National Institute of Standard and
  Technology, FIPS 800-56A, 2018,
  \url{https://doi.org/10.6028/NIST.SP.800-56Ar3}.

\bibitem{FIPS:2023:186-5}
------, \emph{Digital Signature Standard {(DSS)}}, National Institute of
  Standard and Technology, FIPS 186-5, 2023,
  \url{https://doi.org/10.6028/NIST.FIPS.186-5}.

\end{thebibliography}


\end{document}